\documentstyle[amssymb,psfig]{mn}

\title[Mass segregation in young LMC clusters II.]{Mass segregation in
young compact star clusters in the Large Magellanic Cloud: II.  Mass
Functions}

\author[R.  de Grijs et al.]{R.  de Grijs$^1$\thanks{E-mail:
grijs@ast.cam.ac.uk}, G.F.  Gilmore$^1$, R.A.  Johnson$^{1,2}$ and A.D. 
Mackey$^1$\\
$^1$ Institute of Astronomy, University of Cambridge, Madingley Road,
Cambridge CB3 0HA \\
$^2$ European Southern Observatory, Casilla 19001, Santiago 19, Chile
}

\date{Accepted ---. Received ---; in original form ---.}

\pubyear{2001}

\begin{document}
\maketitle

\begin{abstract}
We review the complications involved in the conversion of stellar
luminosities into masses and apply a range of mass-to-luminosity
relations to our {\sl Hubble Space Telescope} observations of the young
LMC star clusters NGC 1805 and NGC 1818.\\
Both the radial dependence of the mass function (MF) and the dependence
of the cluster core radii on mass indicate clear mass segregation in
both clusters at radii $r \lesssim 20-30''$, for masses in excess of
$\sim 1.6-2.5 M_\odot$.  This result does not depend on the mass range
used to fit the slopes or the metallicity assumed. It is clear that the
cluster MFs, at any radius, are not simple power laws.\\
The global and the annular MFs near the core radii appear to be
characterised by similar slopes in the mass range $(-0.15 \le
\log m/M_\odot \le 0.85)$, the MFs beyond $r \gtrsim 30''$ have
significantly steeper slopes. \\
We estimate that while the NGC 1818 cluster core is between $\sim 5$ and
$\sim 30$ crossing times old, the core of NGC 1805 is likely $\lesssim
3-4$ crossing times old.  However, since strong mass segregation is
observed out to $\sim 6 R_{\rm core}$ and $\sim 3 R_{\rm core}$ in NGC
1805 and NGC 1818, respectively, it is most likely that significant
primordial mass segregation was present in both clusters, particularly
in NGC 1805. 
\end{abstract}


\section{Primordial versus Dynamical Mass Segregation}
\label{scenario.sec}

The effects of mass segregation in star clusters, with the more massive
stars being more centrally concentrated than the lower-mass stars,
clearly complicates the interpretation of an observed luminosity
function (LF) at a given position within a star cluster in terms of its
initial mass function (IMF).  Without reliable corrections for the
effects of mass segregation, hence for the structure and dynamical
evolution of the cluster, it is impossible to obtain a realistic global
cluster LF. 

\subsection{Dynamical Evolution in Star Cluster Cores}

Dynamical evolution in dense stellar systems, such as Galactic globular
clusters (GCs) and rich Large Magellanic Cloud (LMC) star clusters,
drives the systems towards energy equipartition, in which the lower-mass
stars will attain higher velocities and therefore occupy larger orbits. 

Consequently, the high-mass stars will gradually sink towards the bottom
of the cluster potential, i.e., the cluster centre (cf.  Spitzer \& Hart
1971), with the highest-mass stars and those closest to the cluster
centre sinking the fastest, although this process is not negligible even
at the cluster's edge (e.g., Chernoff \& Weinberg 1990, Hunter et al. 
1995).  This leads to a more centrally concentrated high-mass component
compared to the lower-mass stellar population, and thus to dynamical
mass segregation.

The time-scale for the onset of significant dynamical mass segregation
is comparable to the cluster's dynamical relaxation time (Spitzer \&
Shull 1975, Inagaki \& Saslaw 1985, Bonnell \& Davies 1998, Elson et al. 
1998).  A cluster's characteristic time-scale is may be taken to be its
half-mass (or median) relaxation time, i.e., the relaxation time at the
mean density for the inner half of the cluster mass for cluster stars
with stellar velocity dispersions characteristic for the cluster as a
whole (Spitzer \& Hart 1971, Lightman \& Shapiro 1978, Meylan 1987,
Malumuth \& Heap 1994, Brandl et al.  1996), and can be written as
(Meylan 1987):
\begin{equation}
\label{trelax.eq}
t_{\rm r,h} = {(8.92 \times 10^5)} {M_{\rm tot}^{1/2} \over {\langle m
\rangle}} {R_{\rm h}^{3/2} \over {\log (0.4 \; M_{\rm tot} / \langle m
\rangle})} {\rm yr,}
\end{equation}
where $R_{\rm h}$ is the half-mass (median) radius (in pc), $M_{\rm
tot}$ the total cluster mass, and $\langle m \rangle$ the typical mass
of a cluster star (both masses in $M_\odot$). 

Although the half-mass relaxation time characterises the dynamical
evolution of a cluster as a whole, significant differences are expected
locally within the cluster.  From Eq.  (\ref{trelax.eq}) it follows
immediately that the relaxation time-scale will be shorter for
higher-mass stars (greater $\langle m \rangle$) than for their
lower-mass companions; numerical simulations of realistic clusters
confirm this picture (e.g., Aarseth \& Heggie 1998, see also Hunter et
al.  1995, Kontizas et al.  1998).  From this argument it follows that
dynamical mass segregation will also be most rapid where the local
relaxation time is shortest, i.e., near the cluster centre (cf.  Fischer
et al.  1998, Hillenbrand \& Hartmann 1998).  The relaxation time in the
core can be written as (Meylan 1987):
\begin{equation}
\label{tcore.eq}
t_{\rm r,0} = {(1.55 \times 10^7)} {{v_s R_{\rm core}^2} \over {\langle
m_0 \rangle \log (0.5 \; M_{\rm tot} / \langle m \rangle})} {\rm yr,}
\end{equation}
where $R_{\rm core}$ is the cluster core radius (in pc), $v_s$ (km
s$^{-1}$) the velocity scale, and $\langle m_0 \rangle$ the mean mass
(in $M_\odot$) of all particles in thermal equilibrium in the central
parts. 

Thus, significant mass segregation among the most massive stars in the
cluster core occurs on the local, central relaxation time-scale
(comparable to just a few crossing times, cf.  Bonnell \& Davies 1998),
whereas a time-scale $\propto t_{\rm r,h}$ is required to affect a large
fraction of the cluster mass. 

It should be kept in mind, however, that even the concept of a ``local
relaxation time'' is only a general approximation, as dynamical
evolution is a continuing process.  The time-scale for a cluster to lose
all traces of its initial conditions also depends on the smoothness of
its gravitational potential, i.e.  the number of stars (Bonnell \&
Davies 1998: larger clusters are inherently smoother, and therefore mass
segregation is slower than in smaller clusters with a grainier mass
distribution), the degree of equipartition reached (e.g., Hunter et al. 
1995: full global, or even local, equipartition is never reached in a
realistic star cluster, not even among the most massive species), and
the slope of the MF (e.g., Lightman \& Shapiro 1978, Inagaki \& Saslaw
1985, Pryor, Smith \& McClure 1986, Sosin 1997: flatter mass spectra
will speed up the dynamical evolution, whereas steep mass spectra will
tend to a higher degree of equipartition), among others. 

In addition, as the more massive stars move inwards towards the cluster
centre, their dynamical evolution will speed up, and hence the dynamical
relaxation time-scale for a specific massive species is hard to define
properly.  This process will be accelerated if there is no (full)
equipartition (cf.  Inagaki \& Saslaw 1985), thus producing high-density
cores very rapidly, where stellar encounters occur very frequently and
binary formation is thought to be very effective (cf.  Inagaki \& Saslaw
1985, Elson et al.  1987b).  In fact, the presence of binary stars may
accelerate the mass segregation significantly, since two-body encounters
between binaries and between binaries and single stars are very
efficient (e.g., Nemec \& Harris 1987, De Marchi \& Paresce 1996,
Bonnell \& Davies 1998, Elson et al.  1998).  This process will act on
similar (or slightly shorter) time-scales as the conventional dynamical
mass segregation (cf.  Nemec \& Harris 1987, Bonnell \& Davies 1998,
Elson et al.  1998).  In summary, the time-scale for dynamical
relaxation is a strong function of position within a cluster, and varies
with its age. 

\subsection{Primordial Mass Segregation}

Although a cluster will have lost all traces of its initial conditions
on time-scales longer than its characteristic relaxation time, on
shorter time-scales the observed stellar density distribution is likely
the result of dynamical relaxation and of the way that star formation
has taken place.  The process is in fact more complicated, as the
high-mass stars evolve on the same time-scale as the lower mass stars
(cf.  Aarseth 1999).  In order to understand the process of mass
segregation in a cluster in detail, we have to get an idea of the amount
of ``primordial'' mass segregation in the cluster. 

The nature and degree of primordial mass segregation is presumably
determined by the properties of interactions of protostellar material
during the star-forming episode in a cluster.  In the classic picture of
star formation (Shu, Adams \& Lizano 1987), interactions are
unimportant, and mass segregation does not occur.  However, Fischer et
al.  (1998) conclude that their observations of NGC 2157 seem to
indicate the picture in which encounters at the early stages in a
cluster's evolution enhance mass accretion due to the merging of
protostellar clumps until the mass of these clumps exceeds the initial
mass of a star to be formed.  More massive stars are subject to more
mergers, hence accrete even more mass (cf.  Larson 1991, Bonnell et al. 
2001a,b and references therein), and therefore dissipate more kinetic
energy.  In addition, they tend to form near the cluster centre, in the
highest-density region, where the encounter-rate is highest (cf.  Larson
1991, Bonnell et al.  1997, 1998, 2001a,b, Bonnell \& Davies 1998). 
This will lead to an observed position-dependent MF containing more
low-mass stars at larger radii compared to the MF in the cluster centre
(although low-mass stars are still present at small radii).  This
scenario is fully consistent with the idea that more massive stars tend
to form in clumps and lower-mass stars form throughout the cluster
(Hunter et al.  1995, Brandl et al.  1996, and references therein). 

Although it has been claimed that the observed mass segregation in R136,
the central cluster in the large star forming complex 30 Doradus in the
LMC, is likely at least partially primordial (e.g., Malumuth \& Heap
1994, Brandl et al.  1996) its age of $\simeq$ 3--4 Myr is sufficiently
long for at least some dynamical mass segregation, in particular of the
high-mass stars in the core ($r \lesssim 0.5$ pc), to have taken place
(cf.  Malumuth \& Heap 1994, Hunter et al.  1995, Brandl et al.  1996). 
On the other hand, the presence of the high-mass Trapezium stars in the
centre of the very young Orion Nebula Cluster (ONC; $\lesssim 1$ Myr,
equivalent to $\simeq$ 3--5 crossing times; Bonnell \& Davies 1998) is
likely largely due to mass segregation at birth (Bonnell \& Davies 1998,
based on numerical simulations; Hillenbrand \& Hartmann 1998, based on
the appearance of the cluster as non-dynamically relaxed, and references
therein).  Bonnell \& Davies (1998) show convincingly that the massive
stars in the core of the ONC most likely originated within the inner
10--20\% of the cluster. 

Hillenbrand \& Hartmann (1998) argue that the young embedded clusters
NGC 2024 and Monoceros R2 also show evidence for primordial mass
segregation, since the outer regions of these clusters (and of the ONC
as well) are not even one crossing time old. 

\section{The Data}

As part of {\sl Hubble Space Telescope (HST)} programme GO-7307, we
obtained deep {\sl WFPC2} {\it V} and {\it I}-band imaging of 7 rich,
compact star clusters in the LMC, covering a large age range.  In de
Grijs et al.  (2001; Paper I) we presented the observational data for
the two youngest clusters in our sample, NGC 1805 and NGC 1818, and
discussed the dependence of the LFs on radius within each cluster.  We
found clear evidence for luminosity segregation within the inner $\sim
30''$ for both clusters, in the sense that the inner annular LFs showed
a relative overabundance of bright stars with respect to the less
luminous stellar population compared to the outer annular LFs. 

In this paper, we will extend our analysis to the associated MFs and
discuss the implications of our results in terms of the IMF and the star
formation process.  In Section \ref{quantification.sec} we will derive
the MF slopes for both clusters, using a number of mass-to-luminosity
(ML) conversions discussed in Section \ref{mlrelations.sec}.  We will
take care to only include main sequence stars belonging to the clusters
in our final MFs; to do so, we will exclude the field LMC red giant
branch stars from the colour-magnitude diagrams (CMDs), with colours
$(V-I) \ge 0.67$ if they are brighter than $V = 22$ mag (see the CMDs in
Johnson et al.  2001 for comparison).  In fact, as can be seen from
these CMDs, the true structure of the HR diagram above the main-sequence
turn-off is very complex.  Stellar populations of different masses
overlap in colour-magnitude space, so that unambiguous mass
determination from isochrone fits, for the handful to the few dozen
stars populating these areas in each cluster, is highly model-dependent
(cf.  Fig.  11 in Johnson et al.  2001).  In a differential analysis
such as presented in this paper, the uncertainties involved in their
mass determinations are too large and systematic (i.e., model
dependent), so that we cannot include these stars in our analysis. 

Table 1 of Paper I contains the fundamental parameters for our two young
sample clusters.  For the analysis in this paper, however, we need to
justify our choice for the adopted metallicity, age, and cluster mass
in more detail. 

\begin{enumerate}

\item {\bf Cluster metallicities} -- For NGC 1805, metallicity
determinations are scarce.  Johnson et al.  (2001) obtained an estimate
of near solar metallicity, from fits to {\sl HST} CMDs.  The only other
metallicity estimate available for NGC 1805, [Fe/H] $\sim -0.30$
(Meliani et al.  1994) is based on the average metallicity of the young
LMC population and is therefore less certain. 

Abundance estimates for NGC 1818, on the other hand, are readily
available, but exhibit a significant range.  The most recent
determination by Johnson et al.  (2001), based on {\sl HST} CMD fits,
similarly suggests near-solar abundance, [Fe/H] $\approx 0.0$. 
Metallicity determinations based on stellar spectroscopy range from
roughly [Fe/H] $\sim -0.8$ (Meliani et al.  1994, Will et al.  1995,
Oliva \& Origlia 1998) to [Fe/H] $\sim -0.4$ (Jasniewicz \& Th\'evenin
1994, Bonatto et al.  1995; see also Johnson et al.  2001). 

For the purposes of the present paper, we will consider the cases of
[Fe/H] = 0.0 and [Fe/H] $= -0.5$ for both clusters. 

\item {\bf Age estimates} -- Various age estimates exist for both
clusters, which are all roughly consistent with each other, although
based on independent diagnostics.  The age range for NGC 1805 is
approximately bracketed by $\log t ({\rm yr}) = 6.95 - 7.00$ (cf.  Bica
et al.  1990, Barbaro \& Olivi 1991, Santos Jr.  et al.  1995,
Cassatella et al.  1996) and $\log t ({\rm yr}) = 7.6 - 7.7$ (cf. 
Barbaro \& Olivi 1991), the most recent determinations favouring younger
ages.  We will therefore adopt an age for NGC 1805 of $\log t ({\rm yr})
= 7.0^{+0.3}_{-0.1}$. 

Numerous age estimates are available for NGC 1818, on average indicating
a slightly older age for this cluster than for NGC 1805.  Most estimates
bracket the age range between $t \sim 15$ Myr (Bica et al.  1990,
Bonatto et al.  1995, Santos Jr.  et al.  1995, Cassatella et al.  1996)
and $t \gtrsim 65$ Myr (cf.  Barbaro \& Olivi 1991), with the most
recent estimates, based on {\sl HST} CMD fits, favouring an age $t
\simeq 20 - 30$ Myr (e.g., Cassatella et al.  1996, Grebel et al.  1997,
Hunter et al.  1997, van Bever \& Vanbeveren 1997, Fabregat \&
Torrej\'on 2000).  We will therefore adopt an age of $t \simeq 25$ Myr
for NGC 1818, or $\log t ({\rm yr}) = 7.4^{+0.3}_{-0.1}$. 

\item {\bf Cluster masses} -- To obtain mass estimates for both
clusters, we first obtained the total {\it V}-band luminosity for each
cluster based on fits to the surface brightness profiles of our longest
CEN exposures, in order to retain a sufficiently high signal-to-noise
ratio even for the fainter underlying stellar component.  We
subsequently corrected these estimates of $L_{V,\rm tot}$ for the
presence of a large number of saturated stars in the long CEN exposures
(i.e, the observations where we located the cluster centre in the {\sl
WFPC2/PC} chip) by comparison with the short CEN exposures.  Although
even in the short CEN exposures there are some saturated stars (cf. 
Section 3.3 in Paper I), their number is small (12 in NGC 1805 and 18 in
NGC 1818), so that we can get firm lower limits of $\log L_{V,\rm tot}
(L_{V,\odot}) = 4.847$ and 5.388 for NGC 1805 and NGC 1818,
respectively. 

Models of single-burst simple stellar populations (e.g., Bruzual \&
Charlot 1996), which are fairly good approximations of coeval star
clusters, predict mass-to-light (M/L) ratios as a function of age, which
we can use to obtain photometric mass estimates for our clusters.  This
leads to mass estimates of $M_{\rm tot} = 2.8^{+3.0}_{-0.8} \times 10^3
M_\odot$ ($\log M_{\rm tot}/M_\odot = 3.45^{+0.31}_{-0.15}$) for NGC
1805 and $M_{\rm tot} = 2.3^{+1.1}_{-0.3} \times 10^4 M_\odot$, or $\log
M_{\rm tot}/M_\odot = 4.35^{+0.18}_{-0.05}$ for NGC 1818. 

Our mass estimate for NGC 1805 is low compared to the only other
available mass, $M_{\rm tot} = 6 \times 10^3 M_\odot$ (Johnson et al. 
2001), although their mass estimate, based on earlier simple stellar
population models, is close to the upper mass allowed by our $1 \sigma$
uncertainty. 

For NGC 1818, our mass estimate is entirely within the probable range
derived by Elson, Fall \& Freeman (1987), i.e.  $4.1 \le \log M_{\rm
tot}/M_\odot \le 5.7$, depending on the M/L ratio, and Hunter et al.'s
(1997) estimate of $M_{\rm tot} = 3 \times 10^4 M_\odot$ falls
comfortably within our $1 \sigma$ uncertainty.  Chrysovergis et al.'s
(1989) determination of $\log M_{\rm tot}/M_\odot = 4.69$ is outside our
$1 \sigma$ error bar; we speculate that the difference between our two
estimates is due to a combination of the single-mass isotropic King
cluster model used by them, versus our photometric mass determination,
and a different treatment of the background stellar contribution. 

\end{enumerate}

\section{Converting luminosity to mass functions}
\label{mlrelations.sec}

The conversion of an observational LF (which we determined for NGC 1805
and NGC 1818 in Paper I), in a given passband {\it i}, $\phi (M_i)$, to
its associated MF, $\xi (m)$, is not as straightforward as often
assumed.  The differential present-day stellar LF, d{\it N}/d$\phi
(M_i)$, i.e.  the number of stars in the absolute-magnitude interval
$[M_i,M_i+{\rm d}M_i]$, and the differential present-day MF, d{\it
N}/d$\xi(m)$, i.e.  the mass in the corresponding mass interval
$[m,m+{\rm d}m]$, are related through d$N = -\phi(M_i) {\rm d}M_i =
\xi(m) {\rm d}m$ (Kroupa 2000), and therefore
\begin{equation}
\label{MLrelation.eq}
\phi(M_i) = -\xi(m) {{\rm d}m \over {\rm d}M_i} .
\end {equation}

Thus, in order to convert an observational LF into a reliable MF, one
needs to have an accurate knowledge of the appropriate ML -- or
mass--absolute-magnitude -- relation, d{\it m}/d$M_i$.  Empirical ML
relations are hard to come by, and have so far only been obtained for
solar-metallicity stars (e.g., Popper 1980; Andersen 1991; Henry \&
McCarthy 1993, hereafter HM93; Kroupa, Tout \& Gilmore 1993, hereafter
KTG93).  The ML relation is, however, a strong function of the stellar
metallicity, and one needs to include corrections for hidden companion
stars to avoid introducing a systematic bias in the derived MF (e.g.,
KTG93, Kroupa 2000).  For the conversion of the present-day MF to the
IMF, one needs additional corrections for stellar evolution on and off
the main sequence, including corrections for age, mass loss, spread in
metallicity and evolution of rotational angular momentum, or spin (cf. 
Scalo 1986, Kroupa 2000).  Although the ML relation is relatively
well-established for stars more massive than $\sim 0.8 M_\odot$, our
rather limited understanding of the lower-mass, more metal-poor stars,
especially of the boundary conditions between the stellar interior and
their atmospheres, have until recently severely limited the
applicability of reliable ML relations to obtain robust MFs at the
low-mass end. 

\subsection{The mass-luminosity relation down to $\sim 0.4 M_\odot$}

As shown by Eq.  (\ref{MLrelation.eq}), it is in fact the {\it slope} of
the ML relation at a given absolute magnitude that determines the
corresponding mass, which is therefore quite model dependent.  This has
been addressed in detail by, e.g., D'Antona \& Mazzitelli (1983), Kroupa
et al.  (1990, 1993), Elson et al.  (1995) and Kroupa \& Tout (1997). 

The slope of the ML relation varies significantly with absolute
magnitude, or mass.  As shown by Kroupa et al.  (1990, 1993) for
solar-metallicity stars with masses $m \lesssim 1 M_\odot$, it has a
local maximum at $M_V \approx 7$, and reaches a minimum at $M_V \approx
11.5$ (see also Kroupa 2000).  This pronounced minimum corresponds to a
maximum in the present-day LF, while the local maximum at $M_V \approx
7$ corresponds to the Wielen dip in the present-day LF of nearby stars
(e.g., Kroupa et al.  1990, D'Antona \& Mazzitelli 1996, and references
therein). 

The local maximum in the derivative of the ML relation at $M_V \approx
7$ ($m \approx 0.7 M_\odot$) is caused by the increased importance of
the H$^-$ opacity in low-mass stars with decreasing mass (KTG93, Kroupa
\& Tout 1997). 

The ML relation steepens near $M_V = 10$ ($m \sim 0.4-0.5 M_\odot$), due
to the increased importance of H$_2$ formation in the outer shells of
main sequence stars, which in turn leads to core contraction (e.g.,
Chabrier \& Baraffe 1997, Baraffe et al.  1998, Kroupa 2000).

Given the non-linear shape of the ML relation and the small slope at the
low-mass end, any attempt to model the ML relation by either a
polynomial fit or a power-law dependence will yield intrinsically
unreliable MFs (cf.  Elson et al.  1995, Chabrier \& M\'era 1997), in
particular in the low-mass regime.  This model dependence is clearly
illustrated by, e.g., Ferraro et al.  (1997), who compared the MFs for
the GC NGC 6752 derived from a variety of different ML relations at that
time available in the literature. 

\subsection{Age and metallicity dependence and corrections for binarity}
\label{binarity.sec}

The exact shape of the ML relation is sensitive to metallicity;
metallicity changes affect the stellar spectral energy distribution and
therefore the (absolute) magnitude in a given optical passband (cf. 
Brewer et al.  1993).  In fact, it has been argued (cf. Baraffe et al. 
1998) that, although the {\it V}-band ML relation is strongly
metallicity-dependent, the {\it K}-band ML relation is only a very weak
function of metal abundance, yielding similar {\it K}-band fluxes for
$[M/H]=-0.5$ and $[M/H]=0.0$.  Although the ML relation is currently
relatively well-determined for solar-metallicity stars with $m \gtrsim
0.8 M_\odot$, at low metallicities the relation remains very uncertain. 
This is partially due to the lack of an empirical comparison, and to our
still relatively poor understanding of the physical properties of these
stars, although major efforts are currently under way to alleviate this
latter problem (e.g., the recent work by the Lyon group). 

Fortunately, as long as we only consider unevolved main sequence stars,
age effects are negligible and can therefore be ignored (cf.  Brewer et
al.  1993, Ferraro et al.  1997). This applies to the current study for
the stellar mass range considered.

Finally, stellar populations contain in general at least 50 per cent of
multiple systems.  The immediate effect of neglecting a significant
fraction of binary stars in our LF-to-MF conversion will be an
underestimate of the resulting MF slope (cf.  Fischer et al.  1998,
Kroupa 2000).  We will come back to this point in Section
\ref{slopedisc.sec}. 

\subsection{A comparison of mass-luminosity relations}

\subsubsection{Luminosity-to-mass conversion for stars with masses below
$1 M_\odot$}

\begin{figure*}
\psfig{figure=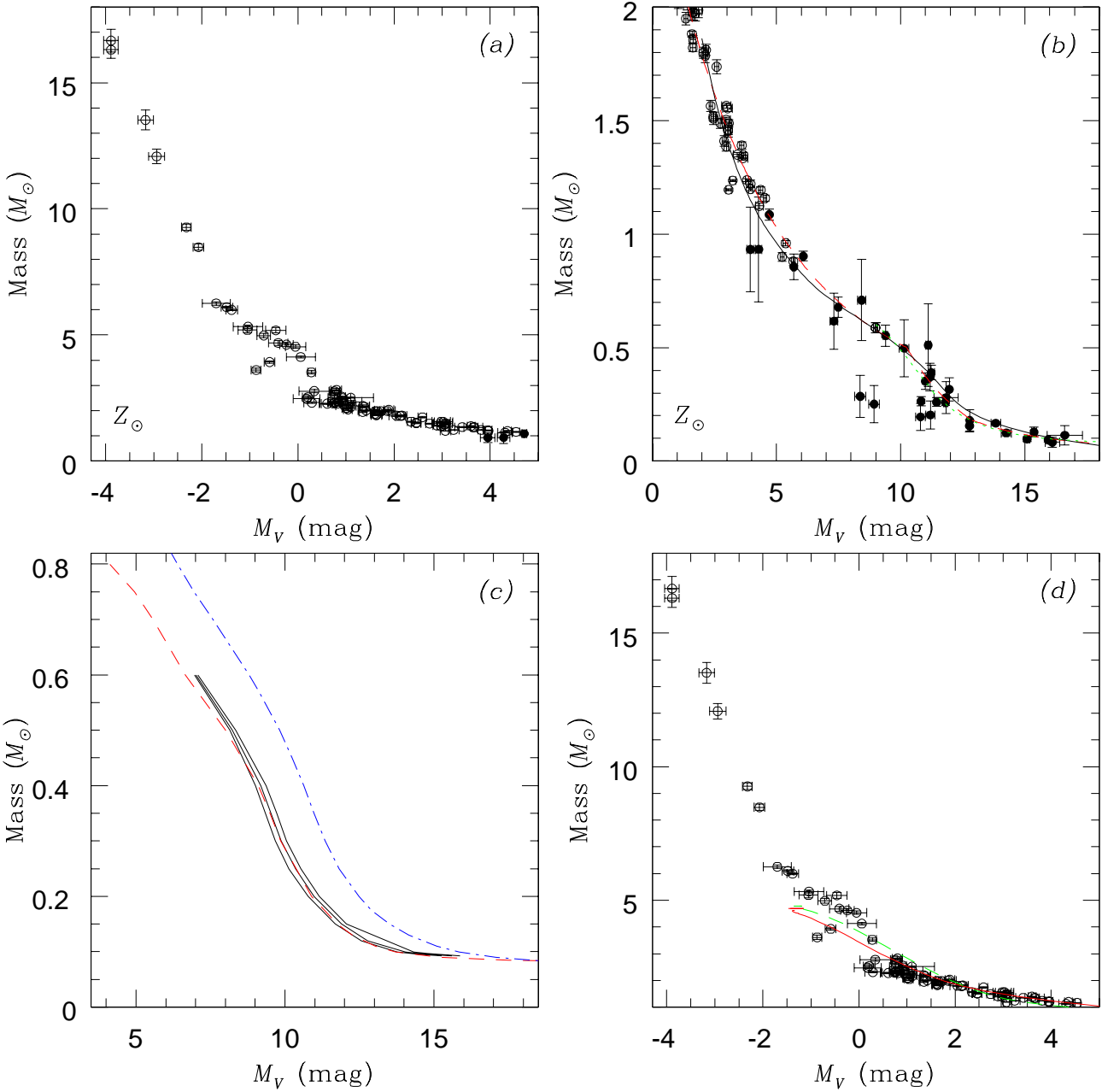,width=18cm}
\caption{\label{MLfig1.fig}Empirical and theoretical ML relations.  {\it
(a)} -- solid bullets: HM93; open circles: Andersen (1991).  {\it (b)}
-- solid line: HM93 fit; dotted line: KTG93 and Kroupa \& Tout (1997)
semi-empirical ML relation; dashed line: Chabrier et al.  (1996)
theoretical ML relation for $m \le 0.6 M_\odot$, based on a third-order
polynomial fit.  {\it (c)} -- Theoretical ML relations for subsolar
abundances: Alexander et al.  (1997; solid lines, for $[M/H] = -1.3,
-1.5, -2.0$ [top to bottom]), and Baraffe et al.  (1997; dashed line,
$[M/H] = -1.5$).  For comparison, the solar-abundance ML relation of
Baraffe et al.  (1997) is also shown (dash-dotted line).  {\it (d)} --
Observational data for $m \ge 1.0 M_\odot$ stars (Andersen 1991), and --
for $m \lesssim 5 M_\odot$ -- theoretical models by GBBC00 for solar
abundance (solid line) and $[M/H] = -1.3$ (dashed line).}
\end{figure*}

Since no empirical ML relations are available for low-mass,
low-metallicity main sequence stars, a test of the goodness of ML
relations in this regime must therefore bear on the comparison of
different models.  Several recent studies have adopted this approach
(e.g., Alexander et al.  1997, Ferraro et al.  1997, Kroupa \& Tout
1997, Piotto, Cool \& King 1997, Saviane et al.  1998). 

For solar-metallicity stars in the mass range $0.1 < m \le 1 M_\odot$,
Leggett et al.  (1996) and Kroupa \& Tout (1997) concluded that,
although {\it all} models considered provided reasonable fits to the
empirical ML relation, the Baraffe et al.  (1995) theoretical ML
relations provided the best overall agreement with all recent
observational constraints.  On the other hand, Bedin et al.  (2001) show
that these are poor at low metallicity.  It should be noted that the
Baraffe et al.  (1995) models were based on {\it grey} model
atmospheres. 

Both Piotto et al.  (1997) and Saviane et al.  (1998), from a comparison
of largely the same theoretical ML relations available in the literature
with observational data for the low-metallicity Galactic GCs NGC 6397
([Fe/H] $\simeq -1.9$) and NGC 1851 ([Fe/H] $\simeq -1.3$),
respectively, concluded that the Alexander et al.  (1997) theoretical ML
relations for the appropriate metallicity provided the best match for
masses $m \lesssim 0.6-0.8 M_\odot$.  Similar conclusions were drawn by
Piotto et al.  (1997) for three other Galactic GCs, M15, M30 and M92. 
Alexander et al.  (1997) themselves found a good to excellent overall
agreement between their models and those of the Lyon group, in
particular the updated Chabrier, Baraffe \& Plez (1996) ones, which
employ the most recent non-grey model atmospheres. 

Figure \ref{MLfig1.fig}a shows the available empirical data, on which
these comparisons are based for solar-metallicity stellar populations. 
The filled bullets represent the HM93 sample; the open circles the
higher-mass Andersen (1991) binary stars.  In panel {\it (b)}, we show
the $m \le 2 M_\odot$ subsample.  Overplotted are the best-fitting
relation of HM93 (solid line), the fit to their semi-empirical ML
relation (dotted line) of KTG93 and Kroupa \& Tout (1997), and the
theoretical ML relation of Chabrier et al.  (1996; dashed line) for
$0.075 \le m \le 0.6 M_\odot$.  The figure shows that the observational
data allow for significant local differences in the slope of the
solar-metallicity ML relation; these uncertainties propagate through the
derivative of the relation when converting LFs to MFs. 

The theoretical ML relation for solar abundance by Chabrier et al. 
(1996) closely follows the most recent semi-empirical ML relation
compiled by Kroupa (KTG93, Kroupa \& Tout 1997).  In Fig. 
\ref{MLfig1.fig}c, we compare the current theoretical ML relations for
subsolar metallicity: the solid lines represent the Alexander et al. 
(1997) ML relations for (top to bottom) $[M/H] \simeq -1.3, -1.5,$ and
$-2.0$\footnote{We applied the procedure outlined in Ryan \& Norris
(1991) to convert the Alexander et al.  (1997) metallicities (Z) to
$[M/H]$ values: $[M/H] \approx [O/H] =$ [O/Fe] + [Fe/H], with [O/Fe] =
+0.35 for [Fe/H] $\le -1$, and [O/Fe] = $-0.35 \times$ [Fe/H] for $-1 <$
[Fe/H] $\le 0$.  Therefore, a stellar population with an observed [Fe/H]
$= -1.5$ corresponds to a model with [Z] = log( Z / Z$_\odot$ ) = $[M/H]
\sim -1.15$, and Z $\simeq 1.35 \times 10^{-3}$.  This procedure has
been shown to apply to halo subdwarfs; these oxygen-enriched abundances
are characteristic of old stellar populations in the Milky Way (cf. 
Baraffe et al.  1995).  Gilmore \& Wyse (1991) have shown that the
element ratios in the LMC are significantly different, however.}; for
reasons of clarity, we only show the $[M/H] = -1.5$ ML relation of
Baraffe et al.  (1997), but the spread due to metallicity differences is
similar to that shown by the Alexander et al.  (1997) relations.  The
most significant differences between both sets of models are seen at
masses $m \gtrsim 0.4 M_\odot$.  This is likely due to the slightly
different treatment of the stellar atmospheres and radiative opacities. 
Finally, for comparison we also show the solar-metallicity theoretical
ML relation of Chabrier et al.  (1996) and Baraffe et al.  (1997). 

\begin{figure*}
\psfig{figure=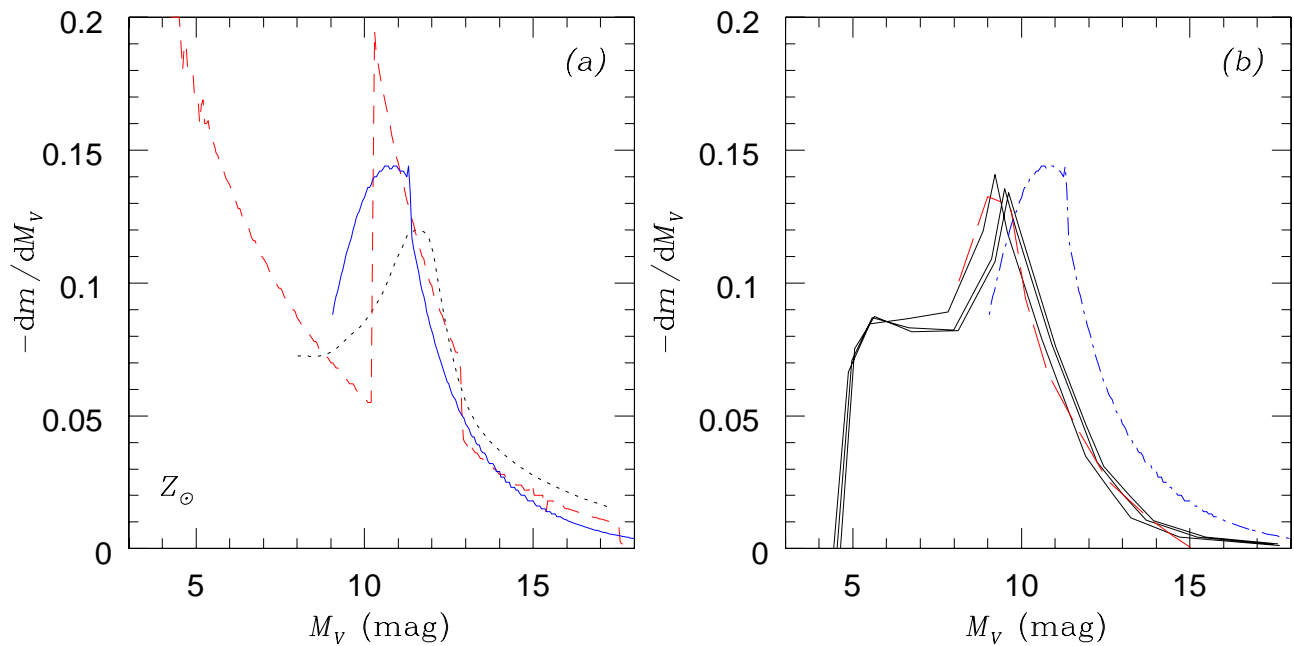,width=18cm}
\vspace{-8.5cm}
\caption{\label{MLfig2.fig}The slope of the ML relation.  {\it (a)} --
Solar-metallicity ML relations: HM93 (dashed line, empirical), KTG93 and
Kroupa \& Tout (1997) (dotted line, semi-empirical), and Chabrier et al. 
(1996) and Baraffe et al.  (1997) (solid line, theoretical).  {\it (b)}
-- Theoretical ML relations for subsolar abundances: Alexander et al. 
(1997; solid lines, for $[M/H] = -1.3, -1.5, -2.0$ [peaking from right
to left]), and Baraffe et al.  (1997; dashed line, $[M/H] = -1.5$).  For
comparison, the solar-abundance ML relation of Chabrier et al.  (1996)
and Baraffe et al.  (1997) is also shown (dash-dotted line, as in {\it
(a)}).}
\end{figure*}

We plot the derivatives of the ML relations as a function of absolute
visual magnitude in Fig.  \ref{MLfig2.fig}.  From Fig. 
\ref{MLfig2.fig}a it is immediately clear that the empirical fit to the
HM93 ML relation inherently leads to unreliable luminosity-to-mass
conversions because of the two sharp discontinuities in the slope. 

Figure \ref{MLfig2.fig}b shows the metallicity dependence of the slope
of the ML relation; the solid lines represent the Alexander et al. 
(1997) ML relations with $[M/H] = -1.3, -1.5,$ and $-2.0$, peaking from
right to left.  The Baraffe et al.  (1997) models (cf.  the dashed line,
for $[M/H] = -1.5$) closely follow the Alexander et al.  (1997) ones. 
For comparison, we have also included the solar-abundance model of
Chabrier et al.  (1996) and Baraffe et al.  (1997), as in panel (a). 

\subsubsection{The more massive stellar population}

The main uncertainties for the luminosity evolution of stars with masses
$m \gtrsim 0.8 M_\odot$ are in the treatment of the degree of mass loss
and convective core overshooting.  Girardi et al.  (2000, hereafter
GBBC00) and Girardi (2001, priv.  comm.) computed a grid of stellar
evolutionary models for stars in the mass range $0.15 \le m \le 7
M_\odot$ for metallicities between $1 \over 50$ and 1.5 times solar,
using updated input physics, as well as moderate core overshooting. 

In Fig.  \ref{MLfig1.fig}d, we show the observational ML relation of
Andersen (1991) for stars with masses $m \ge 1.0 M_\odot$.  In addition,
we have plotted GBBC00's (2000) theoretical ML relations for solar
metallicity (solid line) and for $[M/H] = -1.3$ (dashed line) for stars
less massive than $\sim 5 M_\odot$.  These models include moderate core
overshooting, but the presence or absence of this process in the models
does not significantly change the resulting ML relation for this mass
range. 

\subsubsection{Comparison for our young LMC clusters}

Based on the comparison and discussion in the previous sections, for the
conversion of our observational (individual) stellar magnitudes (Paper I) to
masses, and thence to MFs we will use 
\begin{itemize}

\item the empirical ML relation of HM93.  This ML relation is defined
for stars with $M_V \ge 1.45$. 

\item the KTG93 and Kroupa \& Tout (1997) semi-empirical ML relation for
stars with $M_V \ge 2.00$, with an extension to $M_V = -3$ by adoption
of Scalo's (1986) mass-$M_V$ relation. 

\item the parametrisation of these by Tout et al.  (1996, hereafter
TPEH96), valid for masses in the range $-1 \le \log m/M_\odot \le 2$; we
converted the corresponding bolometric luminosities to absolute {\it
V}-band magnitudes using the bolometric corrections of Lejeune,
Cuisinier \& Buser (1998).  The TPEH96 ML relations are given as a
function of metallicity from $Z = 10^{-4}$ to $Z = 0.03$. 

\item the GBBC00 models.  For solar metallicities, the models for their
youngest isochrone of 60 Myr are defined for stars with $-3.381 \le M_V
\le 12.911$, while for the subsolar abundance of $Z = 0.008$ this
corresponds to $-4.832 \le M_V \le 12.562$. 

\end{itemize}

Although we argued that the Baraffe et al.  (1998, hereafter BCAH98)
models employ the most recent input physics, their mass range, $m
\lesssim 1.0 M_\odot$, precludes us from using their models, since
completeness generally drops below our 50\% limit for $m \lesssim 0.8
M_\odot$, thus leaving us with too few data points for a useful
comparison. 

\begin{figure*}
\psfig{figure=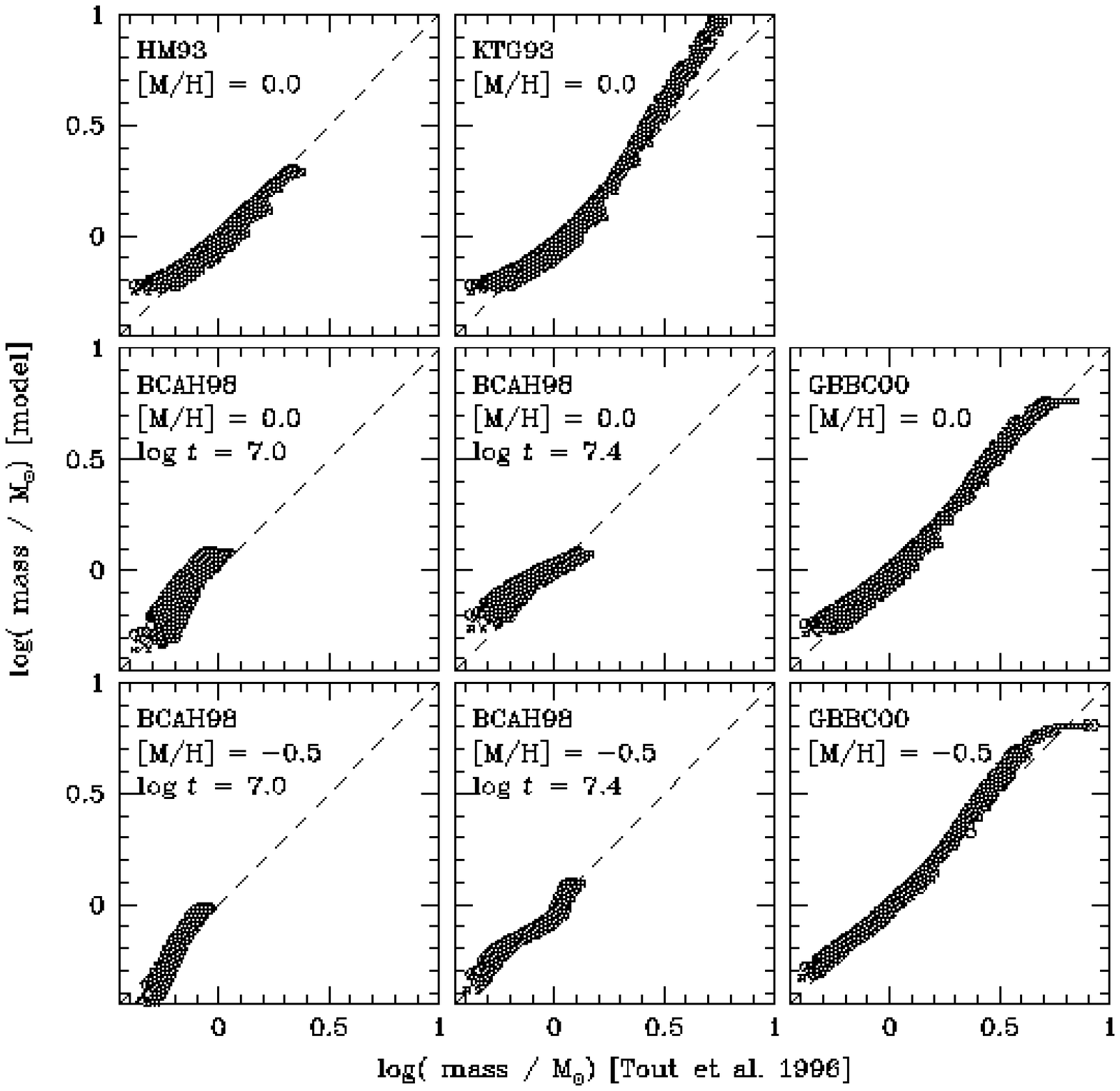,width=18cm}
\caption{\label{compmass.fig}Comparison of the mass estimates for NGC
1805 (open circles) and NGC 1818 (filled circles) resulting from various
ML relations.  We used TPEH96's parametrisation as comparison ML
relation because of its large mass range.}
\end{figure*}

In Fig.  \ref{compmass.fig} we compare the mass estimates based on HM93,
KTG93, BCAH98 and GBBC00 to TPEH96's parametrisation, for both solar and
subsolar ([Fe/H] $=-0.5$) metallicities and ages of $\sim 10$ and 25
Myr.  Significant differences are appreciated among the individual
models, in particular between TPEH96 on the one hand and the high-mass
end ($\log m/M_\odot \gtrsim 0.3$) of KTG93 and between TPEH96 and the
models of BCAH98, which show systematic deviations from the one-to-one
relation indicated by the dashed line.  It is therefore not unlikely
that the differences among the models dominate the uncertainties in the
derived MF and -- ultimately -- in the IMF slope (see also Bedin et al. 
2001, who reached a similar conclusion in their analysis of the Galactic
GC M4).  We will quantify these effects in the next section. 

\begin{figure*}
\psfig{figure=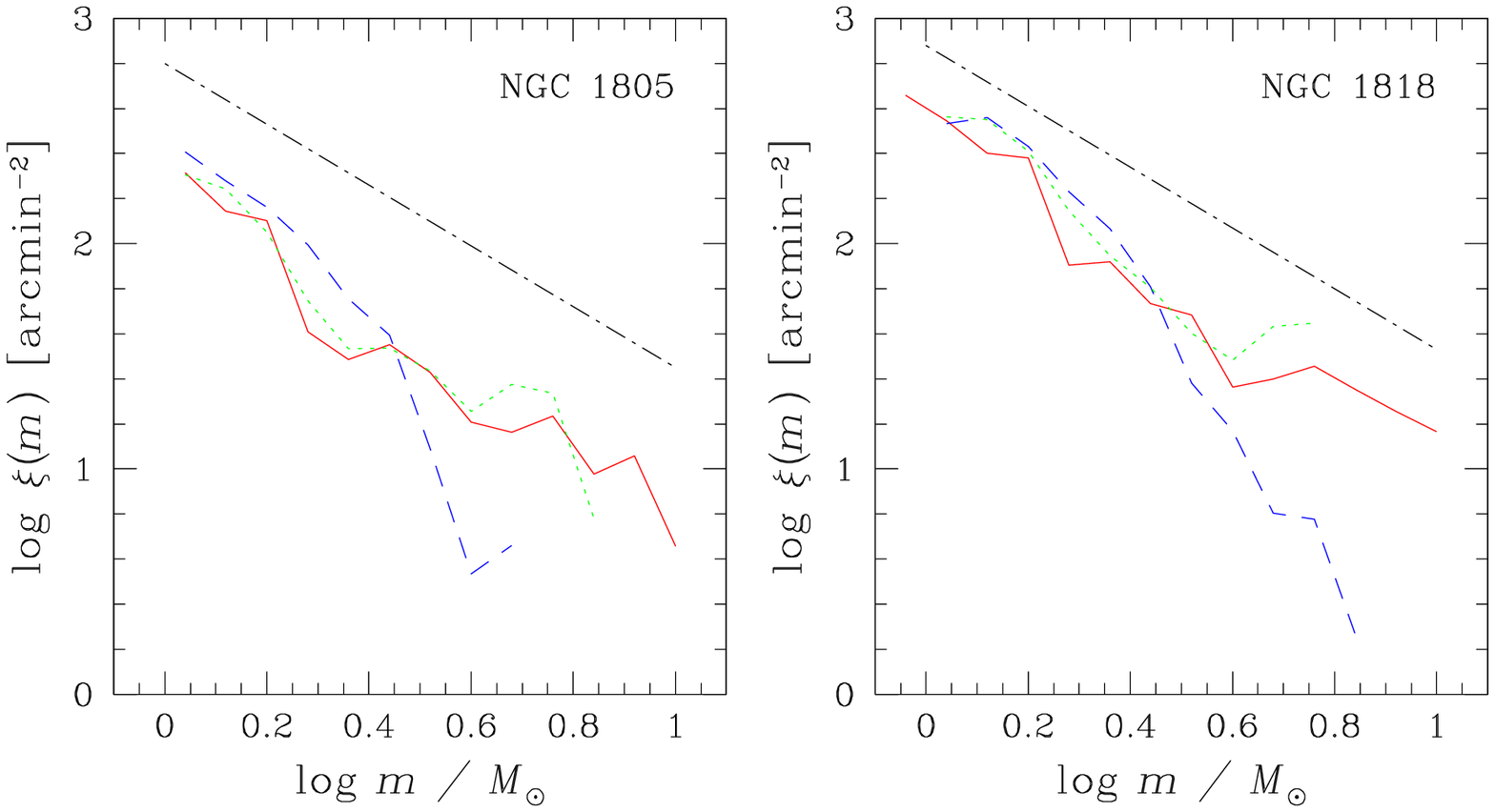,width=18cm}
\vspace*{-8cm}
\caption{\label{figMFs.fig}Examples of the derived global cluster MFs,
normalized to unit area, for three of the ML conversions adopted in this
paper, KTG93 (solid lines), TPEH96 (dashed lines) and GBBC00 (dotted
lines).  All MFs are shown for solar metallicity and for mass bins
corresponding to luminosity ranges that exceed our 50\% completeness
limits (see text).  The dash-dotted lines show the Salpeter IMF
corresponding to the mass range under consideration.}
\end{figure*}

In Fig.  \ref{figMFs.fig} we show examples of the derived global cluster
MFs, normalized to unit area, for three of the ML conversions adopted in
this paper, KTG93, TPEH96 and GBBC00.  All MFs are shown for solar
metallicity and for mass bins corresponding to luminosity ranges that
exceed our 50\% completeness limits (see Paper I).  The MFs, and all
other MFs discussed in this paper, were corrected for incompleteness and
background contamination following identical procedures as for the LFs
in Paper I.  Significant systematic effects result from the adoption of
any given ML conversion, as can be clearly seen.  For comparison, we
also plot a fiducial Salpeter IMF, which appears to be a reasonable
approximation for the global cluster MFs in the mass range $(-0.15
\lesssim \log m/M_\odot \lesssim 0.8)$. 

\section{Quantification of Mass Segregation Effects}
\label{quantification.sec}

\subsection{Radial dependence of luminosity and mass functions}
\label{raddep.sec}

The effects of mass segregation can be quantified using a variety of
methods.  The most popular and straightforward diagnostic for mass
segregation effects is undoubtedly the dependence of MF slope, $\Gamma =
\Delta \log \xi(m) / \Delta m$ (where $\xi(m) \propto m^\Gamma$), on
cluster radius. 

\begin{figure*}
\psfig{figure=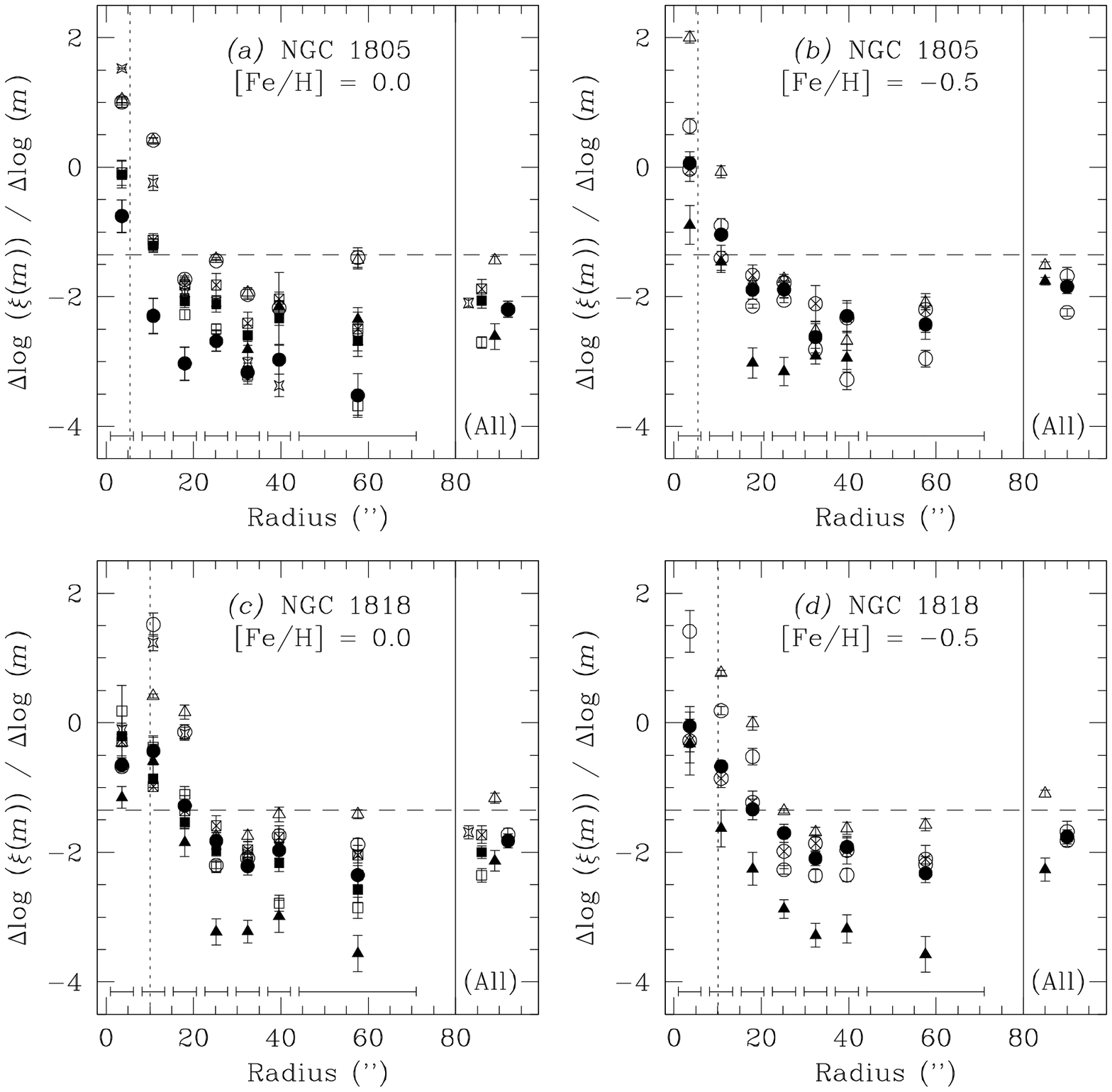,width=18cm}
\caption{\label{mfslopes.fig}Mass function slopes as a function of
cluster radius. The symbols are coded as follows: stars -- HM93, squares
-- KTG93, triangles -- TPEH96, circles -- GBBC00; open symbols
correspond to a mass range $-0.15 \le \log m/M_\odot \le 0.30$, filled
symbols to $-0.15 \le \log m/M_\odot \le 0.70$, and crossed open symbols
extend up to $\log m/M_\odot = 0.85$. The vertical dotted lines indicate
the cluster core radii; the Salpeter slope is represented by
the horizontal dashed lines. The radial ranges over which the MF slopes
were determined are shown by horizontal bars at the bottom of each
panel. The right-hand subpanels show the overall MF slopes for the
clusters as a whole; the data points are spread out radially for display
purposes.}
\end{figure*}

In Fig.  \ref{mfslopes.fig} we plot the derived MF slopes as a function
of cluster radius for our four adopted ML conversions and assuming three
different fitting ranges in mass:
\begin{itemize}
\item $-0.15 \le \log m/M_\odot \le 0.30$ for all conversions;
\item $-0.15 \le \log m/M_\odot \le 0.70$ for KTG93, TPEH96 and GBBC00;
and 
\item $-0.15 \le \log m/M_\odot \le 0.85$ for TPEH96 and GBBC00.
\end{itemize}

The adopted radial ranges for our annular MFs are indicated by the
horizontal bars at the bottom of each panel.  Although we see a large
spread among models and mass fitting ranges, clear mass segregation is
observed in both clusters at radii $r \lesssim 20''$, well outside the
cluster core radii (indicated by the vertical dotted lines).  We have
also indicated the Salpeter (1955) IMF slope, $\Gamma =
-1.35$ (dashed horizontal lines).  While both the global MFs (which are
dominated by the inner, mass-segregated stellar population) and the
annular MFs near the core radii appear to be consistent with the
Salpeter IMF slope (cf.  Vesperini \& Heggie 1997; for the cluster
models assumed here, $R_{\rm core} \approx R_{\rm h}$), the MFs beyond
the cluster radii where mass segregation is significant (i.e., $r
\gtrsim 30''$)are characterised by steeper slopes, i.e., relatively more
low-mass stars compared to high mass stars than found in the inner
cluster regions.  This result holds for all mass fitting ranges, all ML
conversions considered, solar and subsolar ([Fe/H] $= -0.5$)
metallicity, and both clusters.

The error bars in Fig.  \ref{mfslopes.fig} represent the formal
uncertainty in the fits; the systematic uncertainties are clearly
greater, and a strong function both of the adopted ML conversion and of
the mass range used for the fitting of the MF slopes.  In all cases, the
larger mass ranges used for the fitting result in steeper MF slopes than
the smaller (lowest-mass) range, thus presenting clear evidence for
non-power law shaped MFs.  In addition, the TPEH96 parametrisation
results in systematically steeper slopes for all cluster-metallicity
combinations. 

\begin{figure*}
\psfig{figure=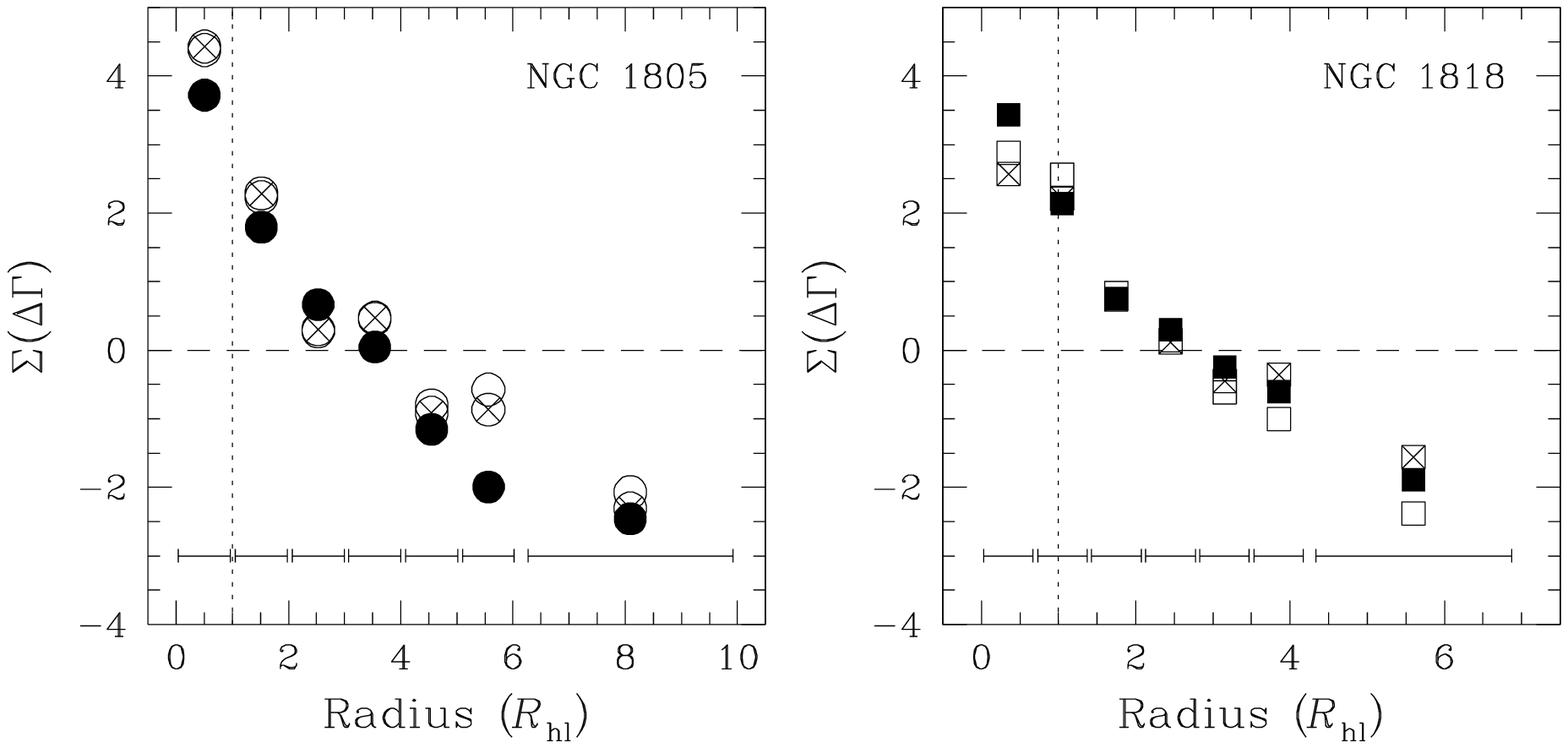,width=18cm}
\vspace*{-9cm}
\caption{\label{robustmf.fig}Deviations of the annular MFs from the
global MF as a function of radius, as discussed in the text.  Filled
symbols were obtained using the ML relation of KTG93, open symbols are
based on TPEH96's ML conversion, and crossed open symbols result from
the GBBC00 models.  The horizontal bars at the bottom of the figure
indicate the radial range used to obtain the data points, the cluster
core radii are indicated by the vertical dotted lines.}
\end{figure*}

Because of the strong model dependence, in particular because of the
sensitivity to the choice of ML relation, and the accuracy of the
corrections for incompleteness and background star contamination of
single power law fits to the annular MFs, in Fig.  \ref{robustmf.fig} we
introduce a more robust characterisation of the presence of mass
segregation in these two young star clusters.  We quantified the
deviations of the high-mass range of the annular MFs from the global MF
following a similar procedure as defined in Paper I for the LFs.  All
annular MFs were normalized to the global MF in the range $0.00 \le \log
m/M_\odot \le 0.20$, where the effects of mass segregation -- if any --
are negligible, as shown above; subsequently, we determined the sum of
the differences between the global and the scaled annular MFs in the
common mass range $0.20 < \log m/M_\odot \le 0.60$, $\Sigma (\Delta
\Gamma)$.  We adopted $\log m/M_\odot = 0.60$ as upper mass limit, so
that we could compare the results of all three ML conversions. 

From Fig.  \ref{robustmf.fig}, it follows that both clusters are mass
segregated within $R \simeq 30''$; beyond, the deviations become
relatively constant with increasing radius.  It is also clear {\it (i)}
that there is no appreciable difference between the strength of the mass
segregation in the two clusters, and {\it (ii)} that the scatter among
the data points from the different models is relatively small.  We
therefore conclude that we have been able to quantify the effects of
mass segregation in a fairly robust way by thus minimising the effects
of the choice of ML conversion. 

It is most likely that the effect referred to as mass segregation is
indeed due to a positional dependence of the ratio of high-mass to
low-mass stars within the clusters, and not to different age
distributions (``age segregation'').  Johnson et al.  (2001) have shown
that for the high-mass stars in both clusters the LF is indeed just a
smoothed (almost) coeval CMD, for which age effects are only of second
order importance.  It is possible that for very low stellar masses, i.e. 
pre-main-sequence stars, age segregation may play a more important role,
but this applies only to stars well below the 50\% completeness limits
for both clusters. 

Finally, we point out that it is generally preferred to use star counts
rather than surface brightness profiles to measure mass segregation
effects (e.g., Elson et al.  1987b, Chernoff \& Weinberg 1990).  Elson
et al.  (1987b) argue that the use of surface brightness profiles by
themselves, although initially used to study mass segregation (e.g., da
Costa 1982, Richer \& Fahlman 1989), is limited in the sense that one
cannot distinguish between these effects and any significant degree of
radial velocity anisotropy in a cluster's outer regions (see also
Chernoff \& Weinberg 1990).  In addition, a comparison between results
obtained from star counts and from surface brightness profiles does not
necessarily trace the same system, since star counts are generally
dominated by main-sequence (and subgiant-branch) stars, while surface
brightness profiles mostly trace the giants (and also subgiant stars) in
a cluster (cf.  Elson et al.  1987b). 

\subsection{Core radii}

Conclusive results on the presence of mass segregation in clusters can
also be obtained by examining the core radii of specific massive stellar
species (e.g., Brandl et al.  1996), or -- inversely -- by measuring the
mean stellar mass within a given radius (e.g., Bonnell \& Davies 1998,
Hillenbrand \& Hartmann 1998).  However, it may not always be feasible
to use this diagnostic, since the individual stellar masses of the
cluster members need to be known accurately, thus providing an
additional observational challenge.  In addition, the results depend
critically on which stars are used to obtain the mean mass, and can be
severely affected by small number statistics (cf.  Bonnell \& Davies
1998). 

\begin{figure*}
\psfig{figure=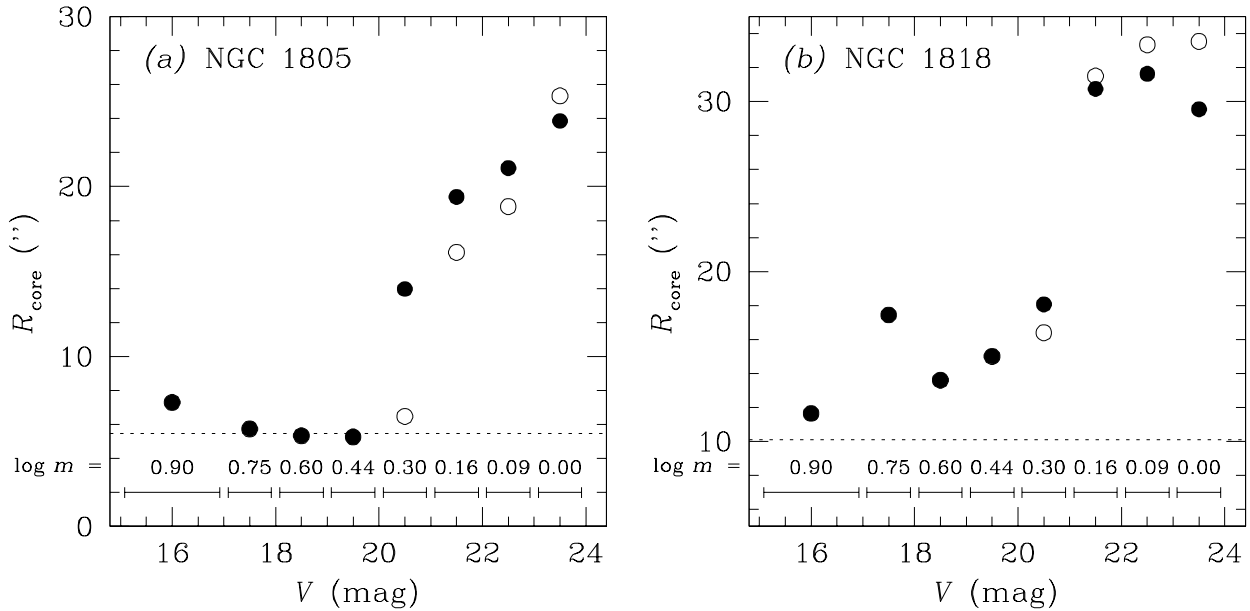,width=18cm}
\vspace{-9cm}
\caption{\label{coreradii.fig}Core radii as a function of magnitude
(mass).  The filled circles are the core radii after correction for the
effects of (in)completeness, area covered by the observations, and
background stars; the open circles are not background subtracted and
serve to indicate the uncertainties due to background correction.  We
have also indicated the mean cluster core radii, obtained from surface
brightness profile fits (dotted lines).  The horizontal bars at the
bottom of the panels indicate the magnitude ranges used to obtain the
core radii; the numbers indicate the approximate mass (in $M_\odot$)
corresponding to the centre of each magnitude range.}
\end{figure*}

\begin{table}
\caption[ ]{\label{coreradii.tab}Cluster core radii as a function of
mass }
{\scriptsize
\begin{center}
\begin{tabular}{ccrccc}
\hline
\hline
\multicolumn{1}{c}{Magnitude} & \multicolumn{1}{c}{$\log m/M_\odot$} &
\multicolumn{2}{c}{NGC 1805} & \multicolumn{2}{c}{NGC 1818}\\
\multicolumn{1}{c}{range $(V)$} & (central) & \multicolumn{1}{c}{$('')$}
& \multicolumn{1}{c}{(pc)} & \multicolumn{1}{c}{$('')$} &
\multicolumn{1}{c}{(pc)} \\
\hline
$15.0 - 17.0$ & 0.90 &  7.30 & 1.85 & 11.65 & 2.95 \\
$17.0 - 18.0$ & 0.75 &  5.74 & 1.45 & 17.45 & 4.41 \\
$18.0 - 19.0$ & 0.60 &  5.34 & 1.35 & 13.61 & 3.44 \\
$19.0 - 20.0$ & 0.44 &  5.27 & 1.33 & 15.01 & 3.80 \\
$20.0 - 21.0$ & 0.30 & 13.97 & 3.53 & 18.07 & 4.57 \\
$21.0 - 22.0$ & 0.16 & 19.39 & 4.91 & 30.74 & 7.78 \\
$22.0 - 23.0$ & 0.09 & 21.09 & 5.34 & 31.63 & 8.00 \\
$23.0 - 24.0$ & 0.00 & 23.86 & 6.04 & 29.55 & 7.48 \\
\\
$15.0 - 24.0$ & & 16.96 & 4.29 & 23.92 & 6.05 \\
\hline
\end{tabular}
\end{center}
{\sc Note:} The range ($15.0 \le V \le 24.0$) represents the cluster
average, corrected for completeness, area covered by the observations,
and the background population.
}
\end{table}
         
Figure \ref{coreradii.fig} and Table \ref{coreradii.tab} show the
dependence of the derived cluster core radius on the adopted magnitude
(or mass) range.  Core radii were derived based on fits to stellar
number counts -- corrected for the effects of incompleteness\footnote{We
only used magnitude (mass) ranges for which the completeness fractions,
as determined in Paper I, were at least 50\%.} and background
contamination (cf.  Paper I) -- of the generalised fitting function
proposed by Elson et al.  (1987a), in the linear regime:
\begin{equation}
\label{elson.eq}
\mu(r) = \mu_0 \Biggl( 1 + \Bigl( {r \over a} \Bigr)^2
\Biggr)^{-\gamma/2} ,
\end{equation} 
where $\mu(r)$ and $\mu_0$ are the radial and central surface
brightness, respectively, $\gamma$ corresponds to the profile slope in
the outer regions of the cluster, and $R_{\rm core} \approx a
(2^{2/\gamma} - 1)^{1/2} \approx R_{\rm h}$.  Equation (\ref{elson.eq})
reduces to a modified Hubble law for $\gamma = 2$, which is a good
approximation to the canonical King model for GCs (King 1966). 

For both NGC 1805 and NGC 1818 we clearly see the effects of mass
segregation for stars with masses $\log m/M_\odot \gtrsim 0.2$ ($M_V
\lesssim 2.4; m \gtrsim 1.6 M_\odot$).  It is also clear that the
brightest four magnitude ranges, i.e.  masses $\log m/M_\odot \gtrsim
0.4 (m \gtrsim 2.5 M_\odot)$, show a similar concentration, while a
trend of increasing core radius with decreasing mass (increasing
magnitude) is apparent for lower masses.  The larger scatter for NGC
1818 is due to the smaller number of stars in each magnitude bin
compared to NGC 1805; for NGC 1818 the associated uncertainties are
determined by a combination of the scatter in the derived core radii and
background effects, while the uncertainties for NGC 1805 are dominated
by the effects of background subtraction. 

We have also indicated the core radii obtained from profile fits to the
overall surface brightness profiles of the clusters.  It is clear that
these are dominated by the mass-segregated high-mass (bright) stars. 

\section{A Comparison of Mass Function Slopes -- Tracing the IMF?}
\label{imf.sec}

\subsection{Comparison with previously published results}

\begin{figure*}
\psfig{figure=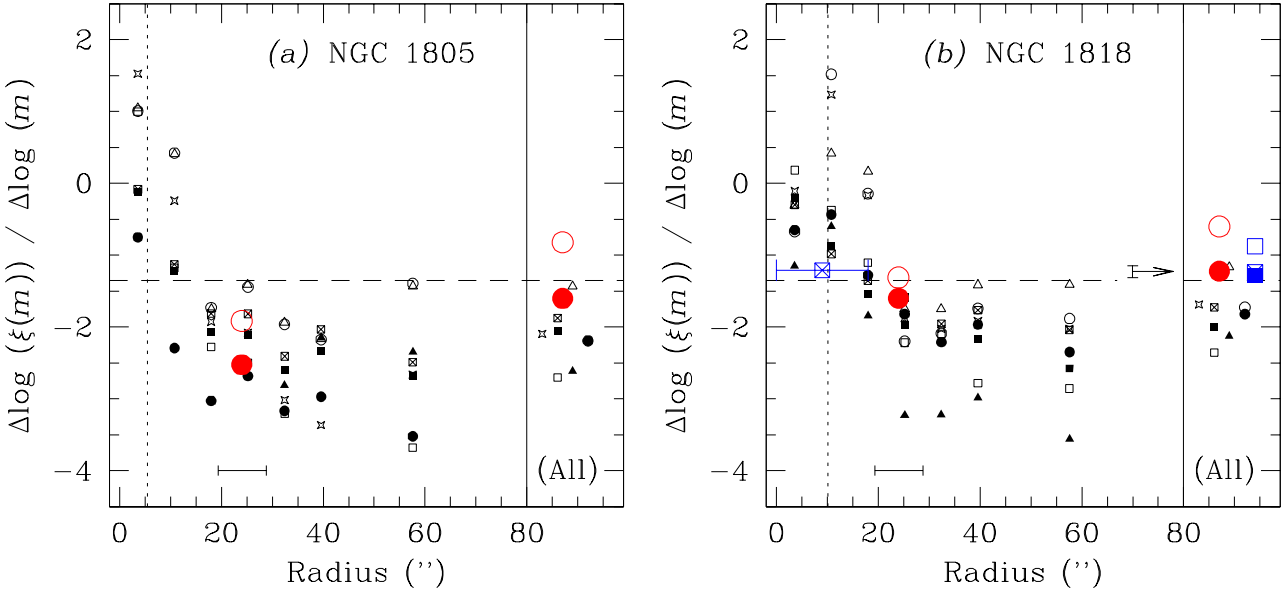,width=18cm}
\vspace{-9cm}
\caption{\label{litslopes.fig}Comparison of MF slopes.  The small
symbols represent our NGC 1805 and NGC 1818 data points; the symbol and
line coding is as in Fig.  \ref{mfslopes.fig}.  The large circles
represent Santiago et al.'s (2001) MF slopes; the large squares those of
Hunter et al.  (1997) for NGC 1818, while the symbol coding is as for
the small symbols.  The radial ranges are again indicated by the
horizontal bars in each panel.  The arrow with vertical error bar in the
panel of NGC 1818 (located at $r = 70''$) shows the global MF slope
derived by Hunter et al.  (1997) and its associated $1 \sigma$
uncertainty.  Solar abundances were assumed.}
\end{figure*}

Few studies have published MFs of sufficient detail and quality for the
two young LMC clusters analysed in this paper to allow useful
comparisons.  Santiago et al.  (2001) published global MFs and MFs
determined in the annulus $4.9 < R < 7.3$ pc ($19.4 < R < 28.8''$) for
both NGC 1805 and NGC 1818, based on the same observations used for the
present study, while Hunter et al.  (1997) published the global MF as
well as the core MF derived from the {\sl HST WFPC2}/PC chip of NGC
1818.  In Fig.  \ref{litslopes.fig} we compare our results with those of
Hunter et al.  (1997) and Santiago et al.  (2001). 

Although Santiago et al.  (2001) used a different ML conversion than
done in the present paper, the slopes and the dependence on the adopted
mass fitting range they derived for their annular MFs are fully
consistent with the range seen at this radius, using any of the ML
relations employed by us.  However, their and Hunter et al.'s (1997)
global MF slopes are somewhat shallower than ours.  The difference is
sufficiently small, however, that it can be explained as due to the
combination of different ML relations and a different treatment of the
background stellar population (see Paper I for a discussion of the
latter).  Hunter et al.  (1997) found no significant difference in MF
slope between the core MF ($\Gamma = -1.21 \pm 0.10$) and the global MF
($\Gamma = -1.25 \pm 0.08$), in the mass range between 0.85 and $9
M_\odot (-0.07 \lesssim \log m/M_\odot \lesssim 0.95)$.  The PC field of
view samples the inner $r \sim 18''$; we have also included their core
data point in Fig.  \ref{litslopes.fig}.  Although Hunter et al.'s
(1997) PC MF slope appears to be slightly steeper than most of our MF
slope determinations at these radii, this can easily be explained as due
to a combination of the uncertainties in the ML conversion used (as
evidenced by the range in MF slopes seen in Fig.  \ref{mfslopes.fig})
and the intrinsic curvature of the MF.  From Fig.  \ref{mfslopes.fig} it
follows that the MF slopes get steeper if increasingly higher mass stars
are included in the fitting range (compare the location of the open,
filled and open-crossed symbols, which indicate increasing mass fitting
ranges).  Hunter et al.'s (1997) PC MF slope determination is based on a
mass range extending up to $9 M_\odot$, while our largest fitting range
only includes stars $\lesssim 7 M_\odot$.  The observed curvature in the
MF will therefore result in a slightly steeper slope for the Hunter et
al.  (1997) MF slope, although still fairly similar to the slopes
determined using our greatest mass fitting ranges at these radii (cf. 
Fig.  \ref{litslopes.fig}).

With regard to the use of different ML relations, it is worth noting
here that the difference between the MF slopes derived by us and those
of both Hunter et al.  (1997) and Santiago et al.  (2001) may be largely
due to the adopted isochrones: while for older clusters the MF slopes
for main-sequence stars are almost independent of age, small differences
between MF slopes as a function of age are appreciated for younger
stellar populations.  This difference is in the sense that using
isochrones for older stellar populations will result in slightly
shallower MF slopes.  This is the most likely explanation for the slight
shift between our MF slopes (based on solar neighbourhood-type stellar
populations) and those of Hunter et al.  (1997) and Santiago et al. 
(2001), who both used younger isochrones to obtain their mass estimates. 
However, the expected steepening in MF slope from evolved to young
stellar populations is almost entirely contained within the observed
spread in MF slope in both clusters (cf.  Hunter et al.  1997).  In
fact, for the ML conversion based on the GBBC00 models, we used their
youngest isochrone (at $t = 6 \times 10^7$ yr); the resulting MFs are
shown as circles in Fig.  \ref{mfslopes.fig}.  Hunter et al.  (1997)
showed that if they had used a 30--40 Myr isochrone instead of the one
at 20 Myr used by these authors, would make the IMF slope appear {\it
shallower} by $\Delta \Gamma \sim 0.15$.  Thus, it appears that the
GBBC00 slopes are entirely consistent with the slopes obtained from the
other ML relations, however, which emphasizes our statement that any
slope difference due to the adoption of different isochrones is
accounted for by the observed spread in MF slopes. 

A further comparison is provided by the result of Will et al.  (1995),
who obtained a MF slope of $\Gamma = -1.1 \pm 0.3$ for stars between 2
and 8 $M_\odot$ (or $\log m/M_\odot = 0.30 - 0.90$).  This slope is
significantly shallower than any of the slopes derived by us, in
particular in view of the observed steepening of the MF slope when
including increasingly higher-mass stars (Section \ref{raddep.sec}).  A
similar discrepancy was already noted by Hunter et al.  (1997), who
argued that this difference (and their greater uncertainties) was most
likely due to the difficulty of obtaining reliable photometry from
crowded ground-based images, while they were not able to resolve stars
in the cluster centre, nor detect stars of similarly faint magnitudes
(low masses) as possible with {\sl HST} observations. 

\subsection{Primordial or dynamical mass segregation?} 

\begin{figure}
\psfig{figure=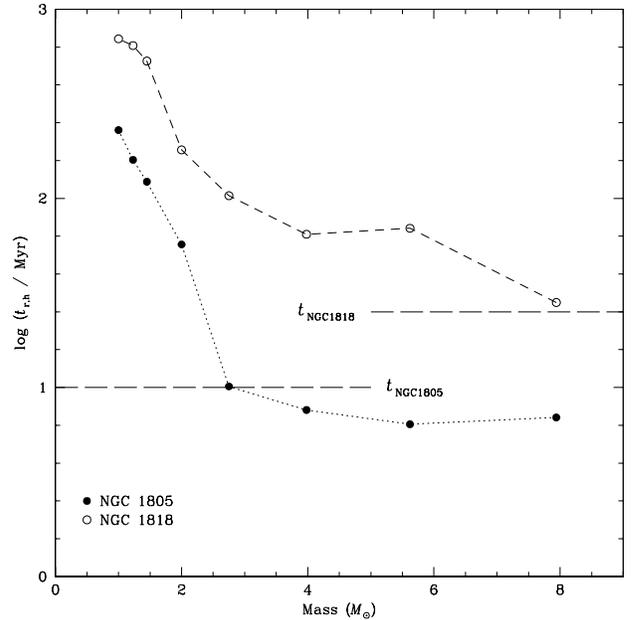,width=9cm}
\caption{\label{trelax.fig}Half-mass relaxation time as a function of
mass for NGC 1805 and NGC 1818. The best age estimates for both clusters
are indicated by horizontal dashed lines.}
\end{figure}

The key question is whether the observed mass segregation in both young
LMC star clusters is the result of the process of star formation itself
or due to dynamical relaxation.  In Fig.  \ref{trelax.fig} we have
plotted the half-mass relaxation time as a function of mass, using Eq. 
(\ref{trelax.eq}) and the mass-dependent core radii of Fig. 
\ref{coreradii.fig}.  For comparison, we have also indicated the ages of
both clusters.  For NGC 1805 significant dynamical mass segregation is
expected to have occurred out to its half-mass radius for stars more
massive than about 3 $M_\odot$ ($\log m/M_\odot \simeq 0.48$), while for
NGC 1818 this corresponds to stars exceeding $\sim 8 M_\odot$ ($\log
m/M_\odot \gtrsim 0.90$). However, from Fig. \ref{coreradii.fig} it
follows that mass segregation becomes significant for masses $m \gtrsim
2.5 M_\odot$, out to at least $20-30''$, or $3-6 R_{\rm core}$.

Dynamical mass segregation in the cluster cores will have occurred on
$10-20 \times$ shorter time-scales, in particular for the more massive
stars (cf.\ Eq.\ (\ref{tcore.eq})).  In fact, if the cluster contains a
significant amount of gas, e.g., $M_{\rm gas} \gtrsim M_{\rm stars}$
(cf.  Lada 1991, Bonnell \& Davies 1998), this will increase the
cluster's gravitational potential, and thus the virialised stellar
velocity dispersion (Bonnell \& Davies 1998).  Therefore, in this case a
larger number of two-body encounters, and hence time, is required to
reach a dynamically relaxed state.  Thus, the relaxation time estimates
obtained by considering only the contributions of the cluster's stellar
component should be considered lower limits, especially for young star
clusters, which are generally rich in gas. 

Elson et al.  (1987b) estimated the central velocity dispersion in NGC
1818 to be in the range $1.1 \lesssim \sigma_0 \lesssim 6.8$ km
s$^{-1}$.  Combining this central velocity dispersion, the core radius
of $\simeq 2.6$ pc, and the cluster age of $\simeq 25$ Myr, we estimate
that the cluster core is between $\sim 5$ and $\sim 30$ crossing times
old, so that dynamical mass segregation in the core should be well under
way.  Although we do not have velocity dispersion information for NGC
1805, it is particularly interesting to extend this analysis to this
younger ($\sim 10$ Myr) cluster.  We know that its core radius is
roughly half that of NGC 1818, and its mass is a factor of $\sim 10$
smaller.  Simple scaling of Eq.  (\ref{trelax.eq}) shows then that the
half-mass relaxation time of NGC 1805 is $\sim 4 - 5\times$ as short as
that of NGC 1818; if we substitute the scaling laws into Eq. 
(\ref{tcore.eq}), we estimate that the central velocity dispersion in
NGC 1805 is $\gtrsim 10\times$ smaller than that in NGC 1818.  From this
argument it follows that the cluster core of NGC 1805 is $\lesssim 3-4$
crossing times old. 

However, since strong mass segregation is observed out to $\sim 6 R_{\rm
core}$ and $\sim 3 R_{\rm core}$ in NGC 1805 and NGC 1818, respectively,
for stellar masses in excess of $\sim 2.5 M_\odot$, it is most likely
that significant primordial mass segregation was present in both
clusters, particularly in NGC 1805. Although this was initially
suggested by Santiago et al.  (2001), we have now substantiated this
claim quantitatively.  Relevant to this discussion is the study by
Bonnell \& Davies (1998), who found that whenever a system of massive
stars is found at the centre of a young star cluster, like the Trapezium
stars in the ONC, a major fraction of it most likely originated in the
inner parts of the cluster.  N-body simulations are currently being
carried out to investigate the fraction of massive stars, and their mass
range, that will have to have originated in the cluster centres to
result in the observed distribution.  We will include these in a
subsequent paper (de Grijs et al., in prep.; Paper III). 

\subsection{The slope of the cluster mass function}
\label{slopedisc.sec}

We will now return to the discussion of Figs.  \ref{mfslopes.fig} and
\ref{litslopes.fig}.  In section \ref{raddep.sec} we showed {\it (i)}
that the slope of the global cluster MF is relatively well approximated
by that at the cluster core radius; {\it (ii)} that at the cluster core
radius the effects of strong mass segregation are still clearly visible;
and {\it (iii)} that in the outer cluster regions, the slope of the
(annular) cluster MFs approaches a constant value. 

Within the uncertainties, we cannot claim that the slopes of the outer
MFs in NGC 1805 and NGC 1818 are significantly different. Starting with
the work by Pryor et al.  (1986) and McClure et al.  (1986), it is
expected that clusters with similar metallicities exhibit similar MF
slopes.  However, Santiago et al.  (2001) claimed to have detected a
significantly different MF slope for both the global and their annular
MF between both clusters.  As is clear from Fig.  \ref{litslopes.fig},
their annular MFs were not taken at sufficiently large radii to avoid
the effects of mass segregation.  Since both clusters are affected by
mass segregation in a slightly different way (which may just be a
reflection of the difference in their dynamical ages), it is not
surprising that Santiago et al.'s (2001) annular MFs exhibit different
slopes. 

Recent studies show that the actual value of the MF slope may vary
substantially from one region to another, depending on parameters such
as the recent star formation rate, metallicity, and mass range (cf. 
Brandl et al.  1996).  The outer cluster regions of R136/30Dor (Malumuth
\& Heap 1994, Brandl et al.  1996), M5 (Richer \& Fahlman 1987), M15
(Sosin \& King 1997) and M30 (Sosin 1997) are all characterized by MF
slopes $\Gamma \lesssim 2.0$.  Bonnell et al.  (2001b) explain this
rather steep MF slope naturally as due to the process of star formation
and accretion itself, resulting from the combination of a gas dominated
and a stellar dominated regime within the forming cluster.  This results
in a double power law IMF, where the lower mass stars have a shallower
slope and the high-mass IMF slope is steeper ($\Gamma \approx -1.5$ and
$-2 \le \Gamma \lesssim -2.5$, respectively), due to the different
accretion physics operating in each regime (i.e., tidal-lobe accretion
versus Bondi-Hoyle accretion for low-mass and high-mass stars, resp.). 

As mentioned in section \ref{binarity.sec}, if there is a significant
percentage of binary or multiple stars in a star cluster, this will lead
us to underestimate the MF slope.  In other words, the MF slopes
determined in this paper are lower limits, because we have assumed that
all stars detected in both clusters are single stars.  Elson et al. 
(1998) found a fraction of $\sim 35 \pm 5$ per cent of roughly similar
mass binaries (with mass of the primary $\sim 2 - 5.5 M_\odot$) in the
centre of NGC 1818, decreasing to $\sim 20 \pm 5$ per cent in the outer
regions of the cluster, which they showed to be consistent with
dynamical mass segregation.  It is not straightforward to correct the
observed LFs for the presence of binaries, in particular since the
binary fraction as a function of brightness is difficult to determine. 
If we consider the information at hand, {\it (i)} the small ($\sim
15$\%) gradient in the total binary fraction derived by Elson et al. 
(1998) for the inner $\sim 3$ core radii of NGC 1818; {\it (ii)} the
result of Rubenstein \& Bailyn (1999) that the binary fraction increases
at fainter magnitudes in the Galactic GC NGC 65752; and {\it (iii)} the
similarity of our annular LFs for faint magnitudes for all annuli and
both clusters, we conclude that the effects of a binary population in
either or both of NGC 1805 and NGC 1818 are likely smaller than the
observed differences in LF shapes as a function of radial distance from
the cluster centres. This is corroborated by Elson et al.'s (1998)
result for NGC 1818.

\section{Summary and Conclusions}

We have reviewed the complications involved in the conversion of
observational LFs into robust MFs, which we have illustrated using a
a number of recently published ML relations.  These ML relations were
subsequently applied to convert the observed LFs of NGC 1805 and NGC
1818, the two youngest star clusters in our {\sl HST} programme of rich
compact LMC star clusters, into MFs. 

The radial dependence of the MF slopes indicate clear mass segregation
in both clusters at radii $r \lesssim 20-30''$, well outside the cluster
core radii.  This result does not depend on the mass range used to fit
the slopes or the metallicity assumed.  In all cases, the larger mass
ranges used for the fitting result in steeper MF slopes than the
smallest mass range dominated by the lowest-mass stars, thus presenting
clear evidence for non-power law shaped MFs.  Within the uncertainties,
we cannot claim that the slopes of the outer MFs in NGC 1805 and NGC
1818 are significantly different.  We also argue that our results are
consistent with previously published results for these clusters if we
properly take the large uncertainties in the conversion of LFs to MFs
into account.  The MF slopes obtained in this paper are in fact lower
limits if there is a significant fraction of binary stars present in the
clusters. 

The global cluster MFs (which are dominated by the inner,
mass-segregated stellar population) and the annular MFs near the core
radii appear to be characterised by similar slopes, the MFs beyond the
cluster radii where mass segregation is significant (i.e., $r \gtrsim
30''$) are characterised by steeper slopes.  It is, however, not unusual
for star clusters to be characterised by rather steep MF slopes; Bonnell
et al.  (2001b) explain this naturally as due to the different accretion
physics operating in the low and high-mass star forming regime. 

We analysed the dependence of the cluster core radius on the adopted
magnitude (mass) range.  For both clusters we clearly detect the effects
of mass segregation for stars with masses $\log m/M_\odot \gtrsim 0.2$
($m \gtrsim 1.6 M_\odot$).  It is also clear that stars with masses
$\log m/M_\odot \gtrsim 0.4 (m \gtrsim 2.5 M_\odot)$ show a similar
concentration, while a trend of increasing core radius with decreasing
mass (increasing magnitude) is apparent for lower masses.  The
characteristic cluster core radii, obtained from profile fits to the
overall surface brightness profiles, are dominated by the
mass-segregated high-mass stars. 

We estimate that the NGC 1818 cluster core is between $\sim 5$ and $\sim
30$ crossing times old, so that dynamical mass segregation in its core
should be well under way.  Although we do not have velocity dispersion
information for NGC 1805, by applying scaling laws we conclude that its
core is likely $\lesssim 3-4$ crossing times old.  However, since strong
mass segregation is observed out to $\sim 6 R_{\rm core}$ and $\sim 3
R_{\rm core}$ in NGC 1805 and NGC 1818, respectively, for stellar masses
in excess of $\sim 2.5 M_\odot$, it is most likely that significant
primordial mass segregation was present in both clusters, particularly
in NGC 1805.  We are currently investigating this further using N-body
simulations. 

\section*{Acknowledgments} We thank Christopher Tout and Mark Wilkinson
for useful discussions and Isabelle Baraffe for making unpublished
subsolar metallicity model results from her NextGen code available to
us.  We acknowledge insightful and constructive comments by the referee. 
This paper is based on observations with the NASA/ESA {\sl Hubble Space
Telescope}, obtained at the Space Telescope Science Institute, which is
operated by the Association of Universities for Research in Astronomy
(AURA), Inc., under NASA contract NAS 5-26555.  This research has made
use of NASA's Astrophysics Data System Abstract Service.


\begin{thebibliography}{}

\bibitem[]{} Aarseth S.J., 1999, PASP, 111, 1333

\bibitem[]{} Aarseth S.J., Heggie D.C., 1998, MNRAS, 297, 794

\bibitem[]{} Alexander D.R., Brocato E., Cassisi S., Castellani V.,
Ciacio F., Degl'Innocenti S., 1997, A\&A, 317, 90

\bibitem[]{} Andersen J., 1991, A\&AR, 3, 91

\bibitem[]{} Bedin L.R., Anderson J., King I.R., Piotto G., 2001, ApJ,
560, L75

\bibitem[]{} Bica E., Alloin D., Santos Jr. J.F.C., 1990, A\&A, 235, 103

\bibitem[]{} Baraffe I., Chabrier G., Allard F., Hauschildt P.H., 1995,
ApJ, 446, L35

\bibitem[]{} Baraffe I., Chabrier G., Allard F., Hauschildt P.H., 1997,
A\&A, 327, 1054

\bibitem[]{} Baraffe I., Chabrier G., Allard F., Hauschildt P.H., 1998,
A\&A, 337, 403 (BCAH98)

\bibitem[]{} Barbaro G., Olivi F.M., 1991, AJ, 101, 922

\bibitem[]{} Bedin L.R., Anderson J., King I.R., Piotto G., 2001, ApJ,
560, L75

\bibitem[]{} Bonatto C., Bica E., Alloin D., 1995, A\&AS, 112, 71

\bibitem[]{} Bonnell I.A., Bate M.R., Clarke C.J., Pringle J.E., 1997,
MNRAS, 285, 201

\bibitem[]{} Bonnell I.A., Bate M.R., Clarke C.J., Pringle J.E., 2001a,
MNRAS, 323, 785

\bibitem[]{} Bonnell I.A., Bate M.R., Zinnecker H., 1998, MNRAS, 298, 93

\bibitem[]{} Bonnell I.A., Clarke C.J., Bate M.R., Pringle J.E., 2001b,
MNRAS, 324, 573

\bibitem[]{} Bonnell I.A., Davies M.B., 1998, MNRAS, 295, 691

\bibitem[]{} Brandl B., Sams B.J., Bertoldi F., Eckart A., Genzel R.,
Drapatz S., Hofmann R., L\"owe M., Quirrenbach A., 1996, ApJ, 466, 254

\bibitem[]{} Brewer J.P., Fahlman G.G., Richer H.B., Searle L., Thompson
I., 1993, AJ, 105, 2158

\bibitem[]{} Bruzual G., Charlot S., 1996, in: Leitherer C., et al.,
1996, PASP, 108, 996 (AAS CDROM Series 7)

\bibitem[]{} Cassatella A., Barbero J., Brocato E., Castellani V., Geyer
E.H., 1996, ApJS, 102, 57

\bibitem[]{} Chabrier G., Baraffe I., 1997, A\&A, 327, 1039

\bibitem[]{} Chabrier G., Baraffe I., Plez B., 1996, ApJ, 459, L91

\bibitem[]{} Chabrier G., M\'era D., 1997, A\&A, 328, 83

\bibitem[]{} Chernoff D.F., Weinberg M.D., 1990, ApJ, 351, 121

\bibitem[]{} Da Costa G.S., 1982, AJ, 87, 990

\bibitem[]{} D'Antona F., Mazzitelli I., 1983, A\&A, 127, 149

\bibitem[]{} D'Antona F., Mazzitelli I., 1996, ApJ, 456, 329

\bibitem[]{} de Grijs R., Johnson R.A., Gilmore G.F., Frayn C.M., 2001,
MNRAS, submitted (Paper I)

\bibitem[]{} De Marchi G., Paresce F., 1996, ApJ, 467, 658

\bibitem[]{} Elson R.A.W., Gilmore G.F., Santiago B.X., Casertano S.,
1995, AJ, 110, 682

\bibitem[]{} Elson R.A.W., Fall S.M., Freeman K.C., 1987a, ApJ, 323, 54

\bibitem[]{} Elson R.A.W., Hut P., Inagaki S., 1987b, ARA\&A, 25, 565

\bibitem[]{} Elson R.A.W., Sigurdsson S., Davies M.B., Hurley J.,
Gilmore G.F., 1998, MNRAS, 300, 857

\bibitem[]{} Fabregat J., Torrej\'on J.M., 2000, A\&A, 357, 451

\bibitem[]{} Ferraro F.R., Carretta E., Bragaglia A., Renzini A.,
Ortolani S., 1997, MNRAS, 286, 1012

\bibitem[]{} Fischer P., Pryor C., Murray S., Mateo M., Richtler T.,
1998, AJ, 115, 592

\bibitem[]{} Gilmore G., Wyse R.F.G., 1991, ApJ, 367, L55

\bibitem[]{} Girardi L., Bressan A., Bertelli G., Chiosi C., 2000,
A\&AS, 141, 371 (GBBC00)

\bibitem[]{} Grebel E.K., 1997, A\&A, 317, 448

\bibitem[]{} Henry T.J., McCarthy D.W., 1993, AJ, 106, 773 (HM93)

\bibitem[]{} Hillenbrand L.A., Hartmann L.E., 1998, ApJ, 492, 540

\bibitem[]{} Hunter D.A., Light R.M., Holtzman J.A., Lynds R., O'Neil Jr.
E.J., Grillmair C.J., 1997, ApJ, 478, 124

\bibitem[]{} Hunter D.A., Shaya E.J., Holtzman J.A., Light R.M., O'Neil
E.J., Lynds R., 1995, ApJ, 448, 179

\bibitem[]{} Inagaki S., Saslaw W.C., 1985, ApJ, 292, 339

\bibitem[]{} Jasniewicz G., Th\'evenin F., 1994, A\&A, 282, 717

\bibitem[]{} Johnson R.A., Beaulieu S.F., Gilmore G.F., Hurley J.,
Santiago B.X., Tanvir N.R., Elson R.A.W., 2001, MNRAS, 324, 367

\bibitem[]{} King I.R., 1966, AJ, 71, 64

\bibitem[]{} Kontizas M., Hatzidimitriou D., Bellas-Velidis I.,
Gouliermis D., Kontizas E., Cannon R.D., 1998, A\&A, 336, 503

\bibitem[]{} Kroupa P., 2000, in: ``Dynamics of Star Clusters and the
Milky Way'', A.S.P.  Conf.  Proc.  228, Deiters S., Fuchs B., Just A.,
Spurzem R., Wielen R., eds., San Francisco: ASP, in press
(astro-ph/0011328; October 2001)

\bibitem[]{} Kroupa P., Tout C.A., 1997, MNRAS, 287, 402

\bibitem[]{} Kroupa P., Tout C.A., Gilmore G.F., 1990, MNRAS, 244, 76

\bibitem[]{} Kroupa P., Tout C.A., Gilmore G.F., 1993, MNRAS, 262, 545
(KTG93) 

\bibitem[]{} Lada C.J., 1991, in: The Physics of Star Formation and
Early Stellar Evolution, Lada C.J., Kylafis N.D., eds., Dordrecht:
Kluwer, p. 329

\bibitem[]{} Larson R.B., 1991, in: Fragmentation of Molecular Clouds
and Star Formation, IAU Symp.  147, Falgarone E., Boulanger F., Duvert
G., eds., Dordrecht: Kluwer, p.  261

\bibitem[]{} Leggett S.K., Allard F., Berriman G., Dahn C.C.,
Hauschildt P.H., 1996, ApJS, 104, 117

\bibitem[]{} Lejeune T., Cuisinier F., Buser R., 1998, A\&AS, 130, 65

\bibitem[]{} Lightman A.P., Shapiro S.L., 1978, Rev. Mod. Phys., 50, 437

\bibitem[]{} Malumuth E.M., Heap S.R., 1994, AJ, 107, 1054

\bibitem[]{} McClure R.D., et al., 1986, ApJ, 307, L49

\bibitem[]{} Meliani M.T., Barbuy B., Richtler T., 1994, A\&A, 290, 753

\bibitem[]{} Meylan G., 1987, A\&A, 184, 144

\bibitem[]{} Meylan G., Heggie D.C., 1997, A\&ARv, 8, 1

\bibitem[]{} Nemec J.M., Harris H.C., 1987, ApJ, 316, 172

\bibitem[]{} Oliva E., Origlia L., 1998, A\&A, 332, 46

\bibitem[]{} Piotto G., Cool A.M., King I.R., 1997, AJ, 113, 1345

\bibitem[]{} Popper D.M., 1980, ARA\&A, 18, 115

\bibitem[]{} Pryor C., Smith G.H., McClure R.D., 1986, AJ, 92, 1358

\bibitem[]{} Richer H.B., Fahlman G.G., 1987, ApJ, 316, 189

\bibitem[]{} Richer H.B., Fahlman G.G., 1989, ApJ, 339, 178

\bibitem[]{} Rubenstein E.P., Bailyn C.D., 1999, ApJ, 513, L33

\bibitem[]{} Ryan S.G., Norris J.E., 1991, AJ, 101, 1865

\bibitem[]{} Salpeter E.E., 1955, ApJ, 121, 161

\bibitem[]{} Santiago B.X., Beaulieu S., Johnson R., Gilmore G.F., 2001,
A\&A, 369, 74

\bibitem[]{} Santos Jr. J.F.C., Bica E., Clar\'\i a J.J., Piatti A.E.,
Girardi L.A., Dottori H., 1995, MNRAS, 276, 1155

\bibitem[]{} Saviane I., Piotto G., Fagotto F., Zaggia S., Capaccioli
M., Aparicio A., 1998, A\&A, 333, 479

\bibitem[]{} Scalo J.M., 1986, Fundam. Cosmic Phys., 11, 1

\bibitem[]{} Shu F.H., Adams F.C., Lizano S., 1987, ARA\&A, 25, 23

\bibitem[]{} Sosin C., 1997, AJ, 114, 1517

\bibitem[]{} Sosin C., King I.R., 1997, AJ, 113, 1328

\bibitem[]{} Spitzer L., 1969, ApJ, 158, L139

\bibitem[]{} Spitzer L., Hart M.H., 1971, ApJ, 164, 399

\bibitem[]{} Spitzer L., Shull J.M., 1975, ApJ, 201, 773

\bibitem[]{} Tout C.A., Pols O.R., Eggleton P.P., Han Z., 1996, MNRAS,
281, 257 (TPEH96)

\bibitem[]{} van Bever J., Vanbeveren D., 1997, A\&A, 322, 116

\bibitem[]{} Vesperini E., Heggie D.C., 1997, MNRAS, 289, 898

\bibitem[]{} Will J.-M., Bomans D.J., de Boer K.S., 1995, A\&A, 295, 54

\end{thebibliography}
\end{document}